\documentclass[preprint,showpacs,preprintnumbers,amsmath,amssymb]{revtex4}

\usepackage{graphicx}

\begin{document}

\title{A new role for adaptive filters in Marchenko equation-based methods for the attenuation of internal multiples}
	
\begin{center}
	\rule{0cm}{2cm}
	{\bf \LARGE A new role for adaptive filters in Marchenko equation-based methods for the attenuation of internal multiples}\\
	\rule{0cm}{1cm}\\
	Myrna Staring, Lele Zhang, Jan Thorbecke and Kees Wapenaar
\end{center}

\hspace{0cm}\textbf{Summary}\\

We have seen many developments in Marchenko equation-based methods for internal multiple attenuation in the past years. Starting from a wave-equation based method that required a smooth velocity model, there are now Marchenko equation-based methods that do not require any model information or user-input. In principle, these methods accurately predict internal multiples. Therefore, the role of the adaptive filter has changed for these methods. Rather than needing an aggressive adaptive filter to compensate for inaccurate internal multiple predictions, only a conservative adaptive filter is needed to compensate for minor amplitude and/or phase errors in the internal multiple predictions caused by imperfect acquisition and preprocessing of the input data. We demonstate that a conservative adaptive filter can be used to improve the attenuation of internal multiples when applying a Marchenko multiple elimination (MME) method to a 2D line of streamer data. In addition, we suggest that an adaptive filter can be used as a feedback mechanism to improve the preprocessing of the input data. 

\pagebreak

\textbf{Introduction.} The use of an adaptive filter for the attenuation of internal multiples is a strongly debated topic. On the one hand, it can be argued that an adaptive filter can damage primary reflections. On the other hand, it is well-known that the field data used for internal multiple prediction have imperfections due to acquisition and preprocessing (e.g., inaccurate deconvolution of the source wavelet), thereby making it very challenging to predict internal multiples with the correct amplitude and phase from field data. Therefore, not using an adaptive filter might result in an incomplete attenuation of the internal multiples and consequently an incorrect interpretation of the target area. 

In recent years, there have been many developments related to Marchenko equation-based methods for internal multiple attenuation \citep{wapenaar2020overview}. Some more conventionally used internal multiple attenuation methods, for example a method proposed by \cite{jakubowicz1998wave}, strongly rely on an adaptive filter to attenuate the internal multiples in field data (the method in principle predicts internal multiples with incorrect amplitudes, uses a layer stripping approach that causes error propagation from shallow to deep, and usually does not remove the source wavelet). In contrast, Marchenko equation-based methods in principle predict internal multiples with the correct amplitude and phase, and thus typically only require a conservative adaptive filter to correct for imperfections due to acquisition and preprocessing. As a result, it is no longer a debate on whether to use an aggressive adaptive filter or no adaptive filter, but on how an adaptive filter can provide a helping hand in ironing out the last details. 

In this paper, we will look at an example of the application of an adaptive Marchenko equation-based method (the Marchenko multiple elimination method) on field data. We will show how only a conservative adaptive filter is needed to improve attenuation of the internal multiples in the data, and how it can also be used as a feedback mechanism.

\textbf{Theory.} The basis of all Marchenko equation-based methods are the coupled Marchenko equations \citep{wapenaar2013three}. By solving these equations iteratively, we retrieve directionally decomposed focusing functions and Green's functions. These wavefields can then be used for different purposes, for example redatuming, internal multiple prediction or homogeneous Green's function retrieval \citep{wapenaar2020overview}. In this paper, we use the projected downgoing focusing function $v^{+}$ as introduced by \cite{van2016adaptive}:

 \vspace*{-5mm}

\begin{equation}
\label{v+}
v^{+}
=
\sum_{k=0}^\infty
(\theta_{t_0}^{t_2} R^{\star} \theta_{t_0}^{t_2}  R)^k
\delta.
\end{equation}

By convolving this function once more with reflection response $R$, we obtain the Marchenko multiple elimination method (MME) \citep{van2016adaptive, zhang2018marchenko}:

\begin{equation}
\label{Rt}
R_t
=
R v^{+}
=
 R \delta
 +
 \alpha
\sum_{k=1}^\infty
R (\theta_{t_0}^{t_2} R^{\star} \theta_{t_0}^{t_2}  R)^k
\delta.
\end{equation}

Internal multiple predictions are obtained by evaluating this series for every timestep $t_2=t-\epsilon$ (where $\epsilon$ represents the band-limitation in the data) and only saving the single sample at timestep $t$. When storing all individual timesteps together, we can add them to the input reflection response to obtain reflection response $R_t$ without internal multiples. A conservative adaptive filter $\alpha$ can be used to adjust the internal multiple predictions when necessary. Although this method is computationally more expensive compared to other Marchenko equation-based methods, it does not require any model information or user-input and is thus completely data-driven. 

\textbf{Example.} \cite{zhangfield} have shown that this method can attenuate internal multiples in streamer data acquired by Equinor in the Norwegian Sea, but perhaps we can do better by using a conservative adaptive filter $\alpha$. The preprocessing of the dataset included 3D to 2D conversion, near offset reconstruction, interpolation to 25 m source and receiver spacing, wavelet deconvolution and the attenuation of surface-related multiples. 

\begin{figure}[!htb]
   \centering
   \includegraphics[width=16.5cm]{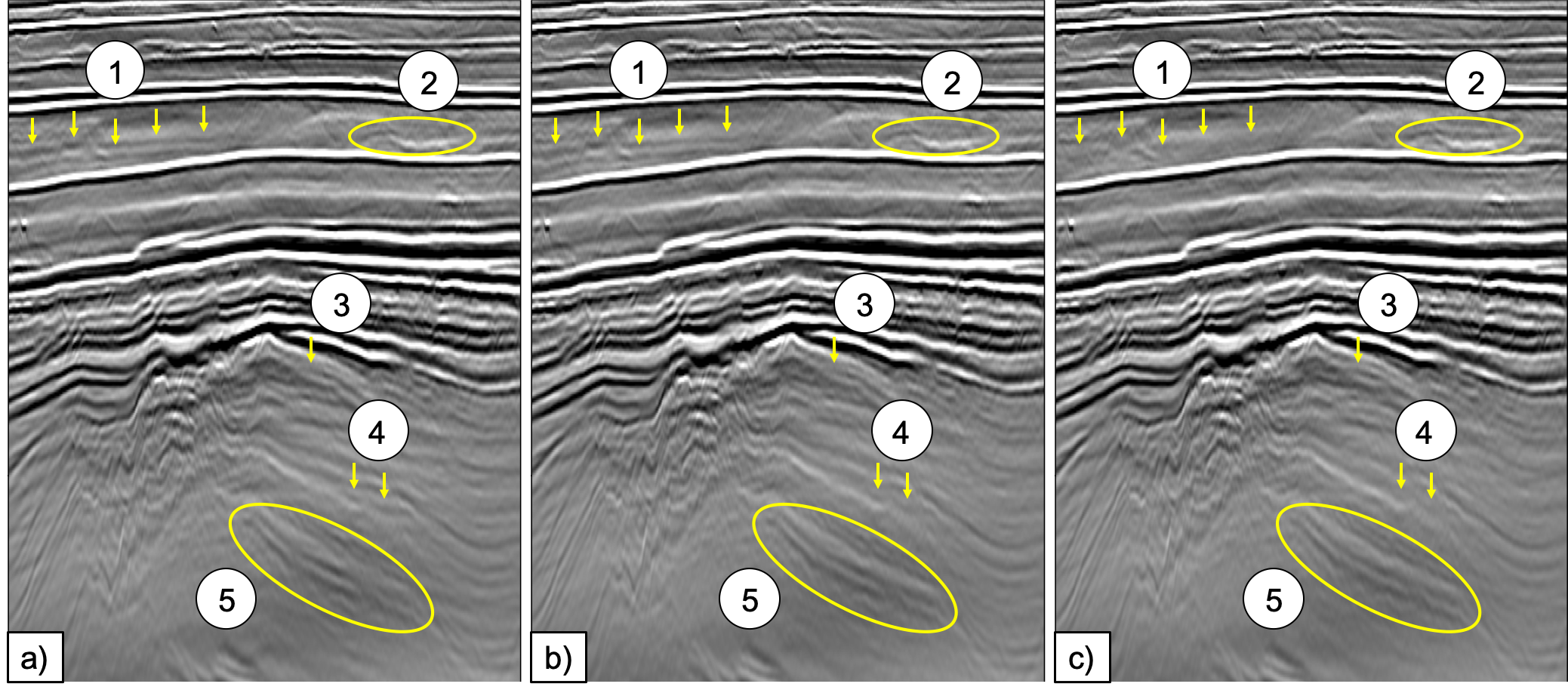}
   \caption{Images of the reflection response a) before internal multiple attenuation, b) after internal multiple attenuation without adaptive filter and c) after internal multiple attenuation using a conservative adaptive filter. \label{fig:1}}
 \end{figure}

Figure \ref{fig:1}a shows an image obtained by one-way wave equation migration of the 2D preprocessed reflection response, while Figure \ref{fig:1}b shows the resulting reflection response after adding 6 terms of the series in Equation \ref{Rt} without adaptive filter (the result presented by \cite{zhangfield}). Instead of simply adding the internal multiple predictions, the result in Figure \ref{fig:1}c was obtained using a conservative adaptive filter $\alpha$ (filter length 3, windows of 200 dt by 50 dx). The arrows at numbers 1 and 4 show a more complete attenuation of what we believe to be internal multiples. The ellipses and the arrow at numbers 2, 3 and 5 show how primary reflections become better visible. In addition, we observed that the adaptive filter mainly changed the amplitudes of the internal multiple predictions, thereby indicating that the input data was not optimally scaled. Based on this observation, we can go back to our preprocessing workflow and optimize the scaling. 

\textbf{Conclusions.} As the prediction of internal multiples is becoming more accurate, there is no longer a need for aggressive adaptive filters. Instead, conservative adaptive filters can be used to attenuate internal multiples in field data more completely. In addition, a conservative adaptive filter can be used as a feedback tool to see whether amplitude and phase of the data were correctly preserved during preprocessing. Still, adaptive filters need to be applied with much care, and a suitable domain for subtraction needs to be chosen for every dataset.

\textbf{Acknowledgements.} We would like to thank Eric Verschuur and Equinor for providing the field data used in this paper. This research was performed in the framework of the project 'Marchenko imaging and monitoring of geophysical reflection data', which is part of the Dutch Open Technology Programme with project number 13939 and is financially supported by NWO Domain Applied and Engineering Sciences. The research of K. Wapenaar has received funding from the European Research Council (ERC) under the European Union's Horizon 2020 research and innovation programme (grant agreement no. 742703). 

\bibliographystyle{apacite}

%
% or
%
%\bibliographystyle{apalike}
%\bibliography{thesis}

\end{document}